\documentclass[11pt,letterpaper]{article}
\usepackage[margin=0.8in]{geometry}
\usepackage[T1]{fontenc}
\usepackage{lmodern}
\usepackage{graphicx,multirow,amsmath,amssymb,amsfonts,booktabs,array,textcomp,bm,algorithm,algorithmic}
\usepackage[numbers,sort&compress]{natbib}
\usepackage[hidelinks]{hyperref}
\usepackage{microtype}
\title{Known Music, Unknown Response: Leakage-Audited Dynamic Arousal Forecasting With Scheduled Future Music}
\author{Bin Wen$^1$ \and Dai-Qiao Zhang$^2$ \and Tien-Ping Tan$^{1,*}$}
\date{}
\begin{document}
\maketitle
\begin{center}\small
$^1$School of Computer Sciences, Universiti Sains Malaysia, 11800, Pulau Pinang, Malaysia\\
$^2$School of Mathematical Sciences, Universiti Sains Malaysia, 11800, Pulau Pinang, Malaysia\\
$^*$Correspondence: \texttt{tienping@usm.my}
\end{center}
\begin{abstract}
Dynamic music emotion recognition usually estimates concurrent or lag-aligned labels from audio that has already been observed. We instead forecast the change in consensus arousal when the upcoming music is scheduled but all later affect labels remain hidden. Evaluation holds out complete songs, encodes each eligible 4-s chunk independently, predicts arousal innovation, and compares the correct future with displaced segments from the same song. On PMEmo and a 1,494-song DEAM evaluation pool excluded from engineering development, a complete past/future MuQ representation raises concordance correlation coefficient (CCC) by 0.389--0.507 over recent affect and remains 0.268--0.382 above a calibrated nonlinear state model. A future-only trajectory contains no past or current audio and still exceeds that state model by 0.150--0.249. The earliest legal future chunk, which ends 5.5--11 s before the target, also remains informative. Correct placement exceeds an overlapping 5-s shift by 0.100--0.209 and the acoustic-nearest eligible same-song segment by 0.078--0.105, with positive song-cluster lower bounds. These results show that both scheduled content and its within-song position carry forecast information under the stated observation rules. The study provides a leakage-controlled reference for known-future music forecasting without treating predictive substitutions as causal interventions.
\end{abstract}


\section{Introduction}
\label{sec:intro}
Music emotion recognition (MER) estimates emotion expressed or perceived in music, frequently in Russell's continuous valence--arousal space \cite{russell1980circumplex}. Dynamic MER extends this objective to a time-varying trajectory. Most systems nevertheless remain recognition models: the acoustic context around an annotated moment is observed, and the model estimates a concurrent or lag-adjusted label \cite{schmidt2011crf,weninger2014online,ma2017multiscale,zhang2023damff}. This article asks a conditional forecasting question instead. If music after a decision time is already scheduled, but its later affective response has not been observed, can the scheduled segment predict a subsequent change in arousal?

The distinction matters in playlist planning, soundtrack adaptation, tutoring media, and other systems where a candidate stimulus is known before its response. Forecasting methods routinely condition on calendar variables or planned interventions whose future values are available at prediction time \cite{lim2021tft}. Music can be treated as such a known future covariate. Doing so does not reveal the future response, but it changes what a model may observe: it may inspect the scheduled waveform while it must not inspect affect labels after the forecast origin.

A concrete use is offline candidate screening. A soundtrack editor could provide a measured group-level starting state and a fixed set of licensed 10--15 s transitions, then use predicted arousal changes to reduce the set requiring human review. This use does not require the model to choose or causally optimize the final track. It does require a population and annotation protocol compatible with the training data, an observable starting state, the same audio-support rules, uncertainty estimates, and human safety and preference checks.

This setting is easy to evaluate incorrectly. Continuous emotion annotations are strongly autocorrelated, so a model can obtain a plausible score by copying the current state. Overlapping windows create many correlated examples from the same song, enabling track or production-style memorization. A bidirectional audio encoder can also cross the declared decision boundary if a long waveform is encoded before features are sliced. Finally, showing that audio improves over state does not establish that the \emph{correctly placed} future segment matters; another segment from the same song may carry nearly the same global style and affect.

\begin{figure*}[t]
\centering
\includegraphics[width=0.98\textwidth]{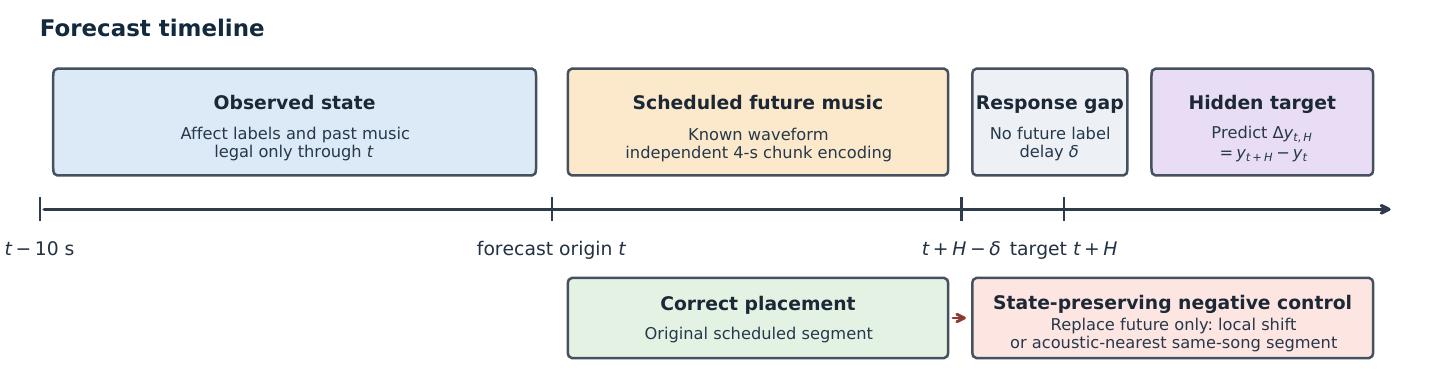}
\caption{Forecast setting. Affect labels after origin $t$ are hidden although the scheduled waveform is known. A valid chunk lies completely inside the permitted future interval. Negative controls preserve the current state and song identity while replacing only the scheduled segment.}
\label{fig:contract}
\end{figure*}

We address these threats with a declared-origin evaluation (Fig.~\ref{fig:contract}). At origin $t$, labels are available only through $t$. The target is the innovation $\Delta y_{t,H}=y_{t+H}-y_t$. Scheduled music may be used only before a response-delay boundary, and every 4-s audio chunk is encoded independently. Outer and inner splits hold out complete songs. Temporal specificity is tested by keeping the state and past anchor fixed while replacing only the future with a local shift or the most acoustically similar eligible segment from the same song.

The primary model combines recent affect with a PCA-256 summary of independently encoded MuQ trajectories \cite{zhu2025muq}. A separate attribution analysis removes all past-audio information and compares a future-only trajectory with a nonlinear state forecaster trained directly on the evaluation target. This identifies scheduled-future information without relying on a weak state baseline or future--past relational descriptors.

The article makes four contributions.
\begin{enumerate}
\item It formulates dynamic arousal forecasting with scheduled music as a known-future-covariate problem with explicit rules for outcomes, audio support, song splits, and temporal controls.
\item It provides an information hierarchy that distinguishes recent affect, observed-past audio, future-only audio, and a complete past/future representation.
\item It introduces state-preserving same-song controls that retain locality or acoustic similarity and therefore test within-song placement rather than merely track identity.
\item It audits audio-support isolation, localization within the scheduled segment, absolute prediction errors, and sensitivity across songs, horizons, and training splits.
\end{enumerate}

Across PMEmo and the engineering-excluded DEAM evaluation pool, the complete representation raises arousal-innovation CCC by 0.389--0.507 over recent affect and by 0.268--0.382 over a calibrated nonlinear state model. Future-only audio remains 0.150--0.249 above that state model. The earliest legal 4-s chunk, ending 5.5--11 s before the target, remains 0.044--0.084 above it; the result is not confined to target-near audio. Correct placement remains 0.078--0.209 above the two hard same-song controls.

\section{Related Work}
\label{sec:related}

\subsection{Dynamic Music Emotion Recognition}
PMEmo and DEAM provide continuous valence--arousal annotations and are widely used for time-varying MER \cite{zhang2018pmemo,aljanaki2017deam}. Earlier systems modeled temporal emotion distributions with conditional random fields or deep recurrent networks \cite{schmidt2011crf,weninger2014online}. Multi-scale context and attention subsequently improved prediction from acoustic neighborhoods \cite{ma2017multiscale,zhang2023damff}, while personalized meta-learning addresses between-listener variation \cite{zhang2025personalized}. ENMI uses past emotion and music features to model emotion flow as multivariate time-series regression \cite{shang2024enmi}.

These systems and the present work solve related but nonidentical tasks. Standard dynamic MER observes acoustic evidence for the moment being labeled and often corrects the response lag. Our setting declares a forecast origin, hides all later annotations, and permits only music that is scheduled and known at that origin. Consequently, published dynamic-MER scores are not direct baselines: their targets, context, splits, and permitted information differ. We compare state and audio forecasts under the same observation rules and use song-disjoint validation.

Table~\ref{tab:taxonomy} makes the task boundary explicit. Our setting is neither ordinary recognition nor causal recommendation. It is a predictive known-covariate problem whose defining object is the information available at the forecast origin.

\begin{table*}[t]
\caption{Task Taxonomy and Information Available at Prediction Time}
\label{tab:taxonomy}
\centering
\footnotesize
\setlength{\tabcolsep}{5.5pt}
\begin{tabular}{p{0.17\textwidth}p{0.16\textwidth}p{0.15\textwidth}p{0.15\textwidth}p{0.20\textwidth}}
\toprule
Task & Target & Audio available & Future response labels & Main identification question \\
\midrule
Static MER & Song/excerpt emotion & Complete target excerpt & Not applicable & What emotion does the excerpt express? \\
Dynamic MER & Concurrent/lagged emotion & Target-era acoustic context & Often used for training alignment & What emotion is expressed now? \\
Past-only affect forecast & Future level/change & History through origin & Hidden & What follows from state and observed history? \\
This work & Future arousal innovation & History plus scheduled future support & Hidden after origin & Does the known future, and its placement, add forecast information? \\
Causal recommendation & Individual outcome under an action & Candidate/action description & Counterfactual unobserved & Which action changes an individual's outcome? \\
\bottomrule
\end{tabular}
\end{table*}

\subsection{Known Future Covariates and Innovation Targets}
Multi-horizon forecasters often separate static, observed, and known-future variables \cite{lim2021tft}. A calendar event, tariff plan, or scheduled control may be known even when the corresponding outcome is not. Treating music as a scheduled covariate places affect forecasting within this framework. Predicting the innovation rather than the future level reduces the direct reward for persistence and makes the target closer to the transition a planning system would need to estimate. It does not remove all state predictability, so state-only models remain necessary controls.

Continuous emotion labels also contain annotation delay and smoothing. Explicit lag correction and joint alignment improve affect modeling \cite{mariooryad2015lag,khorram2021aligning}. We therefore insert a response delay between the permitted music support and the target, with a primary boundary gap of 2 s. Unlike label realignment, this boundary determines which audio chunks are legally observable by the forecasting model.

\subsection{Music Foundation Representations and Interpretable Factors}
MERT and MuQ learn transferable representations from large music collections \cite{li2024mert,zhu2025muq}. Unified and multimodal MER research combines such embeddings with categorical, dimensional, lyrical, or musical features \cite{kang2025music2emo,liyanarachchi2025survey}. Representation quality alone, however, does not show that the correct within-song future is being used. Our negative controls hold the embedding family and song identity fixed while changing temporal placement.

Acoustic studies link loudness, timbre, rhythm, and tonality to multidimensional emotion ratings \cite{eerola2009multidimensional}. Music psychology further associates activation with dynamics, onset strength, temporal regularity, and tempo, whereas valence depends more strongly on harmony, context, familiarity, and semantic appraisal \cite{juslin2008emotional,pring2024social,gu2026interpretable}. Predictive-processing accounts describe listening as hierarchical expectation formation, with sensory and cognitive expectations contributing separately to musical pleasure and surprise \cite{koelsch2019predictive,cheung2023expectations}. These findings motivate studying arousal, while they do not establish the incremental predictive value of a particular hand-crafted factor set.

\subsection{Residual Learning, Cross-Fitting, and Negative Controls}
Residual learning is a standard means of decomposing a forecasting problem, and recent neural objectives explicitly model remaining error streams \cite{liang2026deepboots,wang2026riloss}. Our audio head is fitted to out-of-fold state errors so that its training target does not use an in-sample state prediction. This uses sample separation, as in cross-fitting~\cite{mackey2018orthogonal,liu2022orthogonal}, without introducing a Neyman-orthogonal score.

Negative controls are designed to reveal bias or an unintended pathway by preserving much of the data-generating context while disrupting the relation of interest \cite{lipsitch2010negative}. A random different-song future is too easy for our purpose. We use same-song substitutions and quantify their temporal overlap or acoustic similarity. They are predictive controls, not counterfactual listener outcomes.

\section{Methods and Experimental Setup}
\subsection{Forecast Setting}
\label{sec:problem}

\begin{table}[t]
\caption{Notation and Terms Used in the Forecast Setting}
\label{tab:notation}
\centering
\footnotesize
\setlength{\tabcolsep}{3.0pt}
\begin{tabular}{p{0.23\columnwidth}p{0.68\columnwidth}}
\toprule
Term & Meaning \\
\midrule
$t$ & Forecast origin; labels are observed only through this time \\
$H$ & Forecast horizon \\
$\delta$ & Response gap between legal music support and target \\
$\mathbf s_t$ & Affect history available at the origin \\
$a_{t,H}$ & Scheduled waveform interval permitted at prediction time \\
$\Delta\mathbf y_{t,H}$ & Change from the current to the later affect label \\
Wrong-time control & Same state and song, but a displaced future segment \\
\bottomrule
\end{tabular}
\end{table}

\subsubsection{Target and Permitted Information}
Let $\mathbf y_t\in\mathbb R^2$ be the consensus arousal--valence annotation at time $t$, and let $\mathbf s_t$ contain labels available through $t$. For horizon $H$ and response delay $\delta$, the scheduled music action is
\begin{equation}
a_{t,H}=x_{[t,\,t+H-\delta]},
\label{eq:action}
\end{equation}
where $x$ is the waveform. The target is the change
\begin{equation}
\Delta\mathbf y_{t,H}=\mathbf y_{t+H}-\mathbf y_t.
\label{eq:innovation}
\end{equation}
At inference, $a_{t,H}$ is known but every annotation after $t$ is hidden. We estimate the later level and convert it to an innovation before scoring. The available corpora represent consensus perceived or expressed emotion; the estimand is therefore population-level forecast information, not an individual induced-emotion effect.

We view the response as
\begin{equation}
\mathbf y_{t+H}=m_H(\mathbf s_t)+r_H(a_{t,H},\mathbf s_t)+\boldsymbol\epsilon_{t,H},
\label{eq:decomp}
\end{equation}
where $m_H$ captures state persistence and $r_H$ denotes information associated with the scheduled action. The experiments do not identify $r_H$ causally. They test whether it predicts held-out innovations and whether it changes when the future is displaced within the same song.

\subsubsection{Four Evaluation Rules}
The evaluation imposes four rules.

\textit{Hidden outcomes}: for a held-out prediction, no annotation after its forecast origin enters its input features. Outcomes from outer held-out songs are used only for evaluation, not for fitting, calibration, or model selection. Supervised fitting and inner validation use outcomes only from the corresponding training and development songs.

\textit{Separated audio support}: a frozen music encoder processes each 4-s chunk independently. A chunk is eligible only if its start and end lie entirely inside the declared observed-past or scheduled-future interval. Encoding a whole song and slicing hidden states afterward is prohibited because the encoder could have attended across the boundary.

For example, on an integer-second grid with $t=20$ s, $H=10$ s, and $\delta=2$ s, the target is at 30 s and the legal scheduled support ends at 28 s. With 4-s chunks and 1-s stride, the eligible future chunks are $[20,24],\ldots,[24,28]$; $[25,29]$ is illegal. All labels after 20 s remain hidden. Thus the first legal chunk ends 6 s before the target, whereas the last ends 2 s before it. Half-second annotation origins produce the corresponding 5.5--6.0 s and 2.0--2.5 s gaps used in the experiments.

\textit{Held-out songs}: every outer and inner split holds out complete songs. Standardization, principal component analysis (PCA), calibration and residual targets are fit using training songs only.

\textit{Fixed temporal anchor}: a wrong-time control retains the original affect state, observed-past audio. It replaces only the future segment and recomputes all future-minus-anchor descriptors against the original anchor.

\begin{table}[t]
\caption{Primary Threats and Experimental Responses}
\label{tab:rules}
\centering
\footnotesize
\setlength{\tabcolsep}{3.0pt}
\begin{tabular}{p{0.23\columnwidth}p{0.29\columnwidth}p{0.39\columnwidth}}
\toprule
Threat & Failure mode & Design response \\
\midrule
Emotion inertia & Copy current level & Predict innovation; strong state controls \\
Track identity & Memorize song/style & Hold out complete songs \\
Encoder leakage & Attend beyond support & Independent complete 4-s chunks \\
Coarse content & Any segment works & Same-song local and nearest controls \\
Window dependence & Narrow uncertainty & Song-cluster bootstrap \\
\bottomrule
\end{tabular}
\end{table}

\subsubsection{State-Preserving Temporal Controls}
The unrestricted control chooses an eligible anchor $t'$ from the same song with the same 1-s chunk-sampling phase. Because such a swap can be acoustically easy to distinguish, we use two more demanding label-free controls for the primary placement comparison.

The \emph{local-shift control} chooses the eligible phase-matched candidate closest to a 5-s displacement. At the 15-s horizon it still overlaps the correct permitted interval by approximately 62\%. The \emph{acoustic-nearest control} chooses, among same-song candidates at least 5 s away, the future whose mean MuQ embedding has maximum cosine similarity with the correct candidate. Self-maps are removed before computing margins. We report the realized shift, temporal overlap, cosine similarity, and control coverage so that control difficulty is observable rather than assumed.

The action-over-state and correct-over-control contrasts answer different questions. The former measures the value of scheduled audio. The latter tests within-song placement after style, production, and broad affective content have been preserved. A wrong-time segment can therefore remain better than state only; the relevant test is whether the correctly placed segment is better.

\subsection{Forecasting Models}
\label{sec:method}

\subsubsection{State Model}
The state vector contains the current arousal and valence, differences over one, two, and four annotation steps, a five-frame mean, and elapsed time in seconds. A standardized ridge regressor $f_s$ predicts $\mathbf y_{t+H}$ from $\mathbf s_t$. For the strong-state audit, a fixed histogram gradient booster instead predicts $\Delta y_{t,H}$ directly from 10 s of legal label history. A location--scale map for CCC calibration is estimated solely from inner song-disjoint out-of-fold (OOF) predictions. The outer test songs remain untouched.

\subsubsection{Independently Encoded Music Representation}
MuQ-large encodes every 4-s waveform chunk independently at a 1-s stride. Mean pooling the final hidden-state sequence gives a chunk embedding $\mathbf z_k$. We denote by $\mathbf z_t$ the independently encoded observed-past chunk that ends closest to, and no later than, the forecast origin. For future interval $F$ and matched observed-past interval $P$, the complete relational representation contains
\begin{equation}
\begin{split}
\phi_M={}&[\overline{\mathbf z}_F,\boldsymbol\sigma_F,
\mathbf z_{F,\mathrm{end}}-\mathbf z_t,
\overline{\mathbf z}_F-\overline{\mathbf z}_P,\boldsymbol\eta],\\
\boldsymbol\eta={}&[
\mathrm{mean}_k\|\mathbf z_{k+1}-\mathbf z_k\|_2,
\max_k\|\mathbf z_{k+1}-\mathbf z_k\|_2,
\|\boldsymbol\sigma_F\|_2,\\
&\|\mathbf z_{F,\mathrm{end}}-\mathbf z_t\|_2,
\|\overline{\mathbf z}_F-\overline{\mathbf z}_P\|_2,
 d_c(\overline{\mathbf z}_F,\overline{\mathbf z}_P),\\
&d_c(\mathbf z_{F,\mathrm{end}},\overline{\mathbf z}_P),
 d_c(\overline{\mathbf z}_F,\mathbf z_t),
 d_c(\mathbf z_{F,\mathrm{end}},\mathbf z_t)].
\end{split}
\label{eq:muq}
\end{equation}
Here $\boldsymbol\sigma_F=\mathrm{sd}(\mathbf z_F)$, the path statistics use
adjacent chunks within $F$, and $d_c(\mathbf u,\mathbf v)=1-\cos(\mathbf u,\mathbf v)$.
Thus the representation has four embedding-vector blocks and nine scalars.
The concatenated representation is standardized and projected by whitened
train-fold PCA-256 before ridge regression.
To prevent the action model from simply refitting state, inner-fold OOF state predictions are first produced and the MuQ head is trained on the remaining target error. The base forecast is
\begin{equation}
\widehat{\mathbf y}^{(0)}_{t+H}
=f_s(\mathbf s_t)+f_M(\phi_M(a_{t,H};t)).
\label{eq:base}
\end{equation}

Because $\phi_M$ contains a legal past-audio anchor, its gain over state is described as a \emph{complete past/future representation gain}. Two ablations identify the source of information. Past-only uses a symmetric observed interval and no future descriptor. Future-only uses the future trajectory mean, dispersion, endpoints, and within-future changes; it contains no past or current audio descriptor. Both receive identical PCA-256 capacity and the same OOF state residual target. In the nonlinear-state attribution experiment this target is formed from the raw HGB predictions; CCC calibration is applied only to the separately reported state comparison.

\subsubsection{Training and Inference}
Standardization and PCA are refitted within every relevant training split.
For each outer training set, song-disjoint inner folds generate state
predictions for training the MuQ residual head. The final state model and
MuQ head are then fitted using only the outer training songs. At inference,
the state forecast and audio correction are added, and the observed current
label is subtracted to score innovation. No explicit acoustic-factor
correction is used in the reported forecasts.

\begin{algorithm}[t]
\caption{Outer-Fold State and MuQ Training}
\label{alg:muq}
\begin{algorithmic}[1]
\REQUIRE Training songs $\mathcal D_{tr}$, test songs $\mathcal D_{te}$
\STATE Construct legal state and complete-support MuQ windows
\STATE Cross-fit state predictions within $\mathcal D_{tr}$ using song folds
\STATE Fit train-only standardization, PCA, and MuQ ridge to OOF state residuals
\STATE Fit the state model on all $\mathcal D_{tr}$
\STATE Predict $\mathcal D_{te}$ with the state model plus MuQ head
\STATE Subtract the current label to obtain predicted innovation
\STATE Repeat prediction with state-preserving wrong futures
\end{algorithmic}
\end{algorithm}

After frozen feature extraction, the audio predictor requires only PCA and
ridge regression. The nonlinear state comparison is reported separately
under the same song and audio-support rules.

\subsection{Experimental Design}
\label{sec:experiments}

\subsubsection{Datasets}
\textit{PMEmo} contains 794 chorus clips and dynamic annotations collected from 457 participants at 0.5-s resolution \cite{zhang2018pmemo}. The released dynamic table contains 767 song identifiers. Complete-support filtering retains 611 songs and 19,336 windows at $H=10$ s, and 481 songs and 13,677 windows at $H=15$ s. The principal exclusion is geometric: clips without a complete state, past, future, and response-delay support cannot form a legal example.

\textit{DEAM} contains 1,744 excerpts and 58 full-length songs (1,802 in total) with continuous averaged annotations \cite{aljanaki2017deam}. A partial engineering run exposed results for 305 songs while extraction and support handling were being checked. These songs were not selected by labels or scores, but because their partial outcomes had been observed, all 305 were treated as development data and excluded from primary estimates. On July 10, 2026, before the complete holdout run, a manifest froze their identifiers together with the partial-file hash, method-source hash, response delay, folds, factor subset, shrinkage, and horizons. Of the remaining 1,497 labeled songs, 1,494 yield valid windows: 74,181 at 10 s and 59,241 at 15 s. The complete MuQ archive contains 88,409 independent chunks.

\begin{table}[t]
\caption{Evaluation Data After Complete-Support Filtering}
\label{tab:data}
\centering
\footnotesize
\setlength{\tabcolsep}{4.0pt}
\begin{tabular}{llrrr}
\toprule
Data & Horizon & Songs & Windows & Win./song \\
\midrule
PMEmo & 10 s & 611 & 19,336 & 31.6 \\
PMEmo & 15 s & 481 & 13,677 & 28.4 \\
DEAM holdout & 10 s & 1,494 & 74,181 & 49.7 \\
DEAM holdout & 15 s & 1,494 & 59,241 & 39.7 \\
\bottomrule
\end{tabular}
\end{table}

The archived evaluation rows retain the original intersection of valid
MuQ windows and the timestamp-availability mask used by the earlier factor
pipeline. That mask requires a song trajectory with a frame at or before
the origin and frames in the nominal future interval; songs with fewer
than two retained windows are excluded. We keep this common row set for
all reported comparisons. It is a historical coverage restriction: factor
values are not inputs to the retained state or MuQ predictors, and timestamp
availability does not establish the waveform support of those excluded factors.

Both corpora primarily provide population-average perceived or expressed emotion. We restrict the reported inference to arousal. Valence is not used to support the forecasting claim.

\subsubsection{Splits, Metrics, and Inference}
Every model uses three repeats of five song-disjoint outer folds. Cross-fitted stages use four inner song-disjoint folds. The primary response delay is $\delta=2$ s. All repeated predictions for a window are averaged before inferential resampling.

The primary metric is Lin's concordance correlation coefficient (CCC) on arousal innovation \cite{lin1989ccc}. Mean absolute error (MAE) is secondary. Paired uncertainty is estimated with 5,000 song-cluster bootstrap draws, preserving all correlated windows from a resampled song. Within each corpus, the two horizons form one confirmatory family for the audio-over-state contrast and one for correct-over-basic-wrong-time placement. Each cell uses one-sided $\alpha=0.025$ after correction across the two horizons. Two-sided 97.5\% intervals use bootstrap quantiles 0.0125 and 0.9875; one-sided 97.5\% lower bounds use quantile 0.025. Hard-negative, support-localization, and attribution analyses are diagnostic.

\subsubsection{Baselines and Audits}
The audio baselines predict the same OOF state residual under identical song folds. They add either the mean scheduled MuQ embedding or the complete MuQ representation. The attribution analysis compares state ridge, direct-innovation histogram gradient boosting, past-only audio, future-only audio, and the complete relation with identical PCA-256 heads. The support-localization diagnostic gives the first chunk, last chunk, and complete future mean identical ridge heads under one train-fold PCA-256 basis fitted to a fixed balanced sample of 20,000 representations.

The primary audio head uses PCA-256 and fixed ridge penalties. Attribution and support-localization comparisons match audio-head capacity and keep all preprocessing within training-song folds.

\begin{table}[t]
\caption{Fixed Model and Evaluation Settings}
\label{tab:settings}
\centering
\footnotesize
\setlength{\tabcolsep}{3.2pt}
\begin{tabular}{p{0.37\columnwidth}p{0.53\columnwidth}}
\toprule
Component & Setting \\
\midrule
Audio support & 4-s chunks; 1-s stride \\
State ridge & Standardized; $\alpha=10$ \\
MuQ head & Train-fold PCA-256; ridge $\alpha=100$ \\
Outer/inner splits & $3\times5$ / 4; song-disjoint \\
Response delay & 2 s \\
Bootstrap & 5,000 song-cluster draws \\
\bottomrule
\end{tabular}
\end{table}

Adjacent windows share substantial content, so the number of windows overstates the number of independent musical units. Random cross-validation can be invalid for dependent time-series samples \cite{bergmeir2012cv}; we therefore group every split by song rather than randomly splitting windows. The comparisons are confined to models evaluated under the same observation rules; they do not establish rankings against systems trained or scored with different acoustic context.

\section{Results}
\label{sec:results}

\subsection{Full-Song Encoding Violates the Declared Support}
Before evaluating predictive performance, we test whether encoding an entire song and slicing nominally earlier hidden states respects the declared audio boundary. For three diagnostic PMEmo songs, the waveform length is held fixed and all samples after a cutoff are set to zero. The original and altered waveforms are then separately passed through MuQ, and hidden states with nominal centers before the cutoff are compared. If full-song slicing isolated the permitted context, changing only the suffix would leave those states unchanged.

Table~\ref{tab:encoderleak} shows a material violation. Within the final 10 s before the cutoff, mean relative L2 differences are 0.350--0.445, minimum cosine similarity falls to 0.294--0.366, and individual relative differences exceed 1.0. Even earlier states change. This diagnostic does not estimate how much any particular published score is inflated, but it establishes that post-cutoff waveform information can enter nominally pre-cutoff MuQ states. The tested full-song encode-then-slice procedure therefore fails the declared support-isolation requirement; independent complete-support chunk encoding is the isolation method used here.

\begin{table}[t]
\caption{Suffix-Perturbation Audit of Full-Song MuQ Encoding}
\label{tab:encoderleak}
\centering
\footnotesize
\setlength{\tabcolsep}{3.8pt}
\begin{tabular}{rrrrr}
\toprule
Song & Cutoff & Mean cos. & Min cos. & Rel. L2 \\
\midrule
1 & 19.22 s & 0.923 & 0.366 & 0.350 \\
10 & 19.22 s & 0.890 & 0.307 & 0.445 \\
100 & 24.03 s & 0.899 & 0.294 & 0.429 \\
\bottomrule
\end{tabular}
\end{table}

Values summarize nominally pre-cutoff states in the final 10 s. The audit is intentionally an encoder intervention rather than a model comparison: one nonzero dependence is sufficient to disprove context isolation for full-song slicing, while the reported songs demonstrate that the effect is not confined to one file.

\subsection{Scheduled Music Provides Large Forecast Information}
Table~\ref{tab:action} measures the complete audio representation. It raises arousal-innovation CCC over the state ridge by 0.389--0.507, and every corrected interval is above zero. The future-mean baseline reaches 0.315--0.402 CCC, while the relational trajectory reaches 0.417--0.564. The latter comparison shows that trajectory and anchor-relative structure add information beyond average scheduled content.

\begin{table*}[t]
\caption{Protocol-Matched Forecasts of Arousal Innovation}
\label{tab:action}
\centering
\footnotesize
\setlength{\tabcolsep}{5.5pt}
\begin{tabular}{llrrrr}
\toprule
Data & $H$ & State & MuQ mean & MuQ relation & Relation--state [adj. 97.5\% CI] \\
\midrule
PMEmo & 10 s & 0.028 & 0.315 & 0.417 & +0.389 [0.342, 0.434] \\
PMEmo & 15 s & 0.006 & 0.335 & 0.453 & +0.446 [0.370, 0.514] \\
DEAM holdout & 10 s & 0.041 & 0.317 & 0.458 & +0.416 [0.391, 0.442] \\
DEAM holdout & 15 s & 0.058 & 0.402 & 0.564 & +0.507 [0.475, 0.536] \\
\bottomrule
\end{tabular}
\end{table*}

PMEmo retains fewer songs at 15 s because a longer horizon requires more complete support. To separate this composition change from the horizon comparison, we reevaluated both horizons on the 481 songs eligible at 15 s, without refitting or selecting models. At 10 s the common-song state and complete-representation CCCs are 0.030 and 0.413 (gain 0.383; corrected lower bound 0.337); at 15 s they are 0.006 and 0.453 (gain 0.446; lower bound 0.382). These values closely match Table~\ref{tab:action}, so the difference between horizons is not driven by the additional 130 songs in the 10-s set.

The observed-past audio anchor does not explain the main result. Table~\ref{tab:attribution} replaces the original state ridge with the nonlinear direct-innovation model and separates past-only from future-only audio. The future-only trajectory reaches 0.310--0.429 CCC. It exceeds the calibrated nonlinear state model by 0.150--0.249 and a symmetric past-only trajectory by 0.151--0.261. The complete relation remains 0.268--0.382 above the calibrated nonlinear state model. Corrected lower bounds for future-minus-past range from 0.093 to 0.236. Thus, scheduled-future information survives both stronger state modeling and complete removal of past/current audio descriptors.

\begin{table*}[t]
\caption{Strong-State and Past/Future Attribution. ``HGB raw'' is the direct-innovation output used to form residual targets; ``cal.'' applies an inner-fold location--scale map only for the CCC state comparison.}
\label{tab:attribution}
\centering
\scriptsize
\setlength{\tabcolsep}{3.5pt}
\begin{tabular}{llrrrrrr}
\toprule
Data & $H$ & State ridge & HGB raw/cal. & Past only & Future only & Full relation & Future--past [LB] \\
\midrule
PMEmo & 10 s & 0.028 & 0.075/0.159 & 0.158 & 0.310 & 0.427 & +0.151 [0.112] \\
PMEmo & 15 s & 0.006 & 0.081/0.127 & 0.163 & 0.323 & 0.462 & +0.160 [0.093] \\
DEAM holdout & 10 s & 0.041 & 0.094/0.161 & 0.128 & 0.339 & 0.465 & +0.211 [0.190] \\
DEAM holdout & 15 s & 0.058 & 0.129/0.180 & 0.168 & 0.429 & 0.562 & +0.261 [0.236] \\
\bottomrule
\end{tabular}
\end{table*}

A support-localization analysis tests whether this result is merely recognition from audio close to the target. It fits identical future-only heads to the first legal 4-s chunk, the last legal chunk, or the mean of all legal future chunks; no past/current audio descriptor is present. At $H=10$ s, the first chunk ends 5.5--6.0 s before the target; at $H=15$ s, it ends 10.5--11.0 s before it. Table~\ref{tab:support-location} shows that this early chunk still exceeds the calibrated HGB state model by 0.044--0.084 CCC, with positive corrected lower bounds in all four cells. Later and complete-support representations are stronger, and the full-support mean exceeds the early chunk by 0.076--0.198 CCC. Hence information is already present away from the target and accumulates across the scheduled segment; the main effect cannot be reduced to reading only target-adjacent audio. This diagnostic was not used for model or support selection.

\begin{table*}[t]
\caption{Candidate-Only Localization Within the Legal Future Support (CCC)}
\label{tab:support-location}
\centering
\footnotesize
\setlength{\tabcolsep}{6.0pt}
\begin{tabular}{llrrrrr}
\toprule
Data & $H$ & State HGB (cal.) & Earliest 4 s & Latest 4 s & Full-support mean & Early--state [LB] \\
\midrule
PMEmo & 10 s & 0.159 & 0.243 & 0.285 & 0.319 & +0.084 [0.036] \\
PMEmo & 15 s & 0.127 & 0.205 & 0.306 & 0.339 & +0.078 [0.026] \\
DEAM holdout & 10 s & 0.161 & 0.241 & 0.290 & 0.338 & +0.080 [0.046] \\
DEAM holdout & 15 s & 0.180 & 0.224 & 0.352 & 0.422 & +0.044 [0.005] \\
\bottomrule
\end{tabular}
\end{table*}

The raw HGB predictions generate the held-out residual targets used by the three audio heads; the calibrated HGB column above is a CCC-only state stress test and is not used for fitting. Table~\ref{tab:absolute-errors} therefore reports MAE and RMSE for this same raw-HGB attribution pipeline. The complete relational representation reduces both errors relative to raw HGB in every cell. Its MAE reduction is 3.7--10.0\%, while its RMSE reduction is 10.0--18.2\%. The future-only model reduces RMSE in all four cells and MAE in three; its PMEmo 15-s MAE is 1.2\% higher. Thus, the large CCC lift is not only a location--scale calibration effect; the complete representation also lowers absolute and squared prediction error.

\begin{table*}[t]
\caption{Absolute-Error Audit for the Strong-State Attribution Pipeline}
\label{tab:absolute-errors}
\centering
\footnotesize
\setlength{\tabcolsep}{6.0pt}
\begin{tabular}{llrrrrrr}
\toprule
& & \multicolumn{3}{c}{MAE} & \multicolumn{3}{c}{RMSE} \\
\cmidrule(lr){3-5}\cmidrule(lr){6-8}
Data & $H$ & HGB raw & Future only & Full relation & HGB raw & Future only & Full relation \\
\midrule
PMEmo & 10 s & 0.02531 & 0.02525 & \textbf{0.02438} & 0.03681 & 0.03502 & \textbf{0.03312} \\
PMEmo & 15 s & 0.03174 & 0.03212 & \textbf{0.03032} & 0.04545 & 0.04382 & \textbf{0.04052} \\
DEAM holdout & 10 s & 0.05562 & 0.05497 & \textbf{0.05260} & 0.09283 & 0.08497 & \textbf{0.08004} \\
DEAM holdout & 15 s & 0.07713 & 0.07383 & \textbf{0.06941} & 0.12250 & 0.10816 & \textbf{0.10019} \\
\bottomrule
\end{tabular}
\end{table*}

\begin{figure*}[t]
\centering
\includegraphics[width=0.98\textwidth]{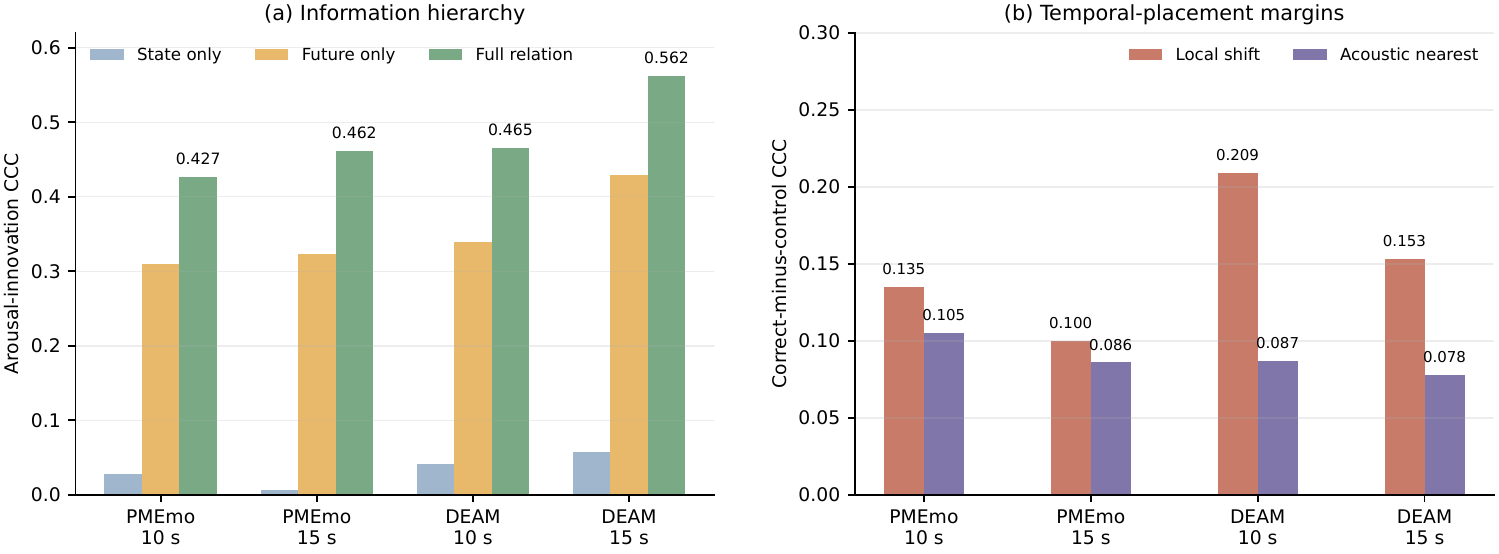}
\caption{Main empirical hierarchy. (a) Candidate-only future audio is well above state, and the complete past/future relation is higher still. (b) Correct placement remains superior after the future is replaced by a strongly overlapping local shift or by the acoustic-nearest eligible same-song segment.}
\label{fig:summary}
\end{figure*}

Figure~\ref{fig:summary}(a) visualizes the information hierarchy. The large effect comes from adding a support-separated scheduled-audio representation to recent affect.

\subsection{Correct Within-Song Placement Matters}
An arbitrary same-song permutation may be an easy negative. Table~\ref{tab:hard} therefore reports the two hard controls. The local candidate is displaced by approximately 5 s and at $H=15$ s retains 0.615--0.629 temporal overlap with the correct permitted interval. Even under this substantial overlap, correct-minus-local margins are 0.100--0.209 CCC. The acoustic-nearest candidate has cosine similarity 0.939--0.979 with the correct future, yet correct-minus-nearest margins remain 0.078--0.105. All diagnostic lower bounds are positive.

\begin{table*}[t]
\caption{Hard Same-Song Temporal Controls}
\label{tab:hard}
\centering
\footnotesize
\setlength{\tabcolsep}{5.0pt}
\begin{tabular}{llrrrrrr}
\toprule
Data & $H$ & Local shift & Overlap & Local margin [LB] & NN cosine & NN margin [LB] \\
\midrule
PMEmo & 10 s & 4.86 s & 0.392 & +0.135 [0.108] & 0.947 & +0.105 [0.078] \\
PMEmo & 15 s & 4.83 s & 0.629 & +0.100 [0.071] & 0.979 & +0.086 [0.063] \\
DEAM holdout & 10 s & 5.00 s & 0.375 & +0.209 [0.185] & 0.939 & +0.087 [0.074] \\
DEAM holdout & 15 s & 5.00 s & 0.615 & +0.153 [0.131] & 0.973 & +0.078 [0.067] \\
\bottomrule
\end{tabular}
\end{table*}

Local self-maps are 0--0.37\%; acoustic-nearest self-maps are 0--8.47\%, and are excluded from their margins. Every local-placement difference is positive in all 60 repeat--fold cells. A marginal standardized-distance diagnostic does not mark the controls as more extreme than the correct inputs, although it is not a joint density test. The unrestricted phase-matched permutation yields larger 0.129--0.237 margins, but the placement conclusion no longer depends on this easier comparison.

The full action lift is larger than the placement margin because a displaced segment from the same song still supplies instrumentation, production, tempo range, and global affective style. The placement comparison asks whether the correct location adds information after those properties are preserved, not whether a wrong segment becomes equivalent to no music.

The placement result is also stable across training seeds and outer folds. Table~\ref{tab:repeat-stability} summarizes the foldwise correct-minus-local diagnostic before predictions are pooled. Every one of the 15 repeat--fold cells is positive for every dataset/horizon pair; repeat-level standard deviations are 0.0014--0.0043 CCC. These are dispersion diagnostics rather than independent tests, while the song-cluster intervals in Table~\ref{tab:hard} remain the inferential results.

\begin{table}[t]
\caption{Repeat/Fold Stability of the Local-Placement Margin}
\label{tab:repeat-stability}
\centering
\scriptsize
\setlength{\tabcolsep}{3.5pt}
\begin{tabular}{llrrr}
\toprule
Data & $H$ & Repeat SD & Fold range & Positive \\
\midrule
PMEmo & 10 s & 0.0043 & [0.086, 0.192] & 15/15 \\
PMEmo & 15 s & 0.0035 & [0.075, 0.144] & 15/15 \\
DEAM & 10 s & 0.0026 & [0.152, 0.241] & 15/15 \\
DEAM & 15 s & 0.0014 & [0.089, 0.175] & 15/15 \\
\bottomrule
\end{tabular}
\end{table}

\subsection{The Audio Representation Is Most Useful for Large Changes}
\begin{figure*}[t]
\centering
\includegraphics[width=0.98\textwidth]{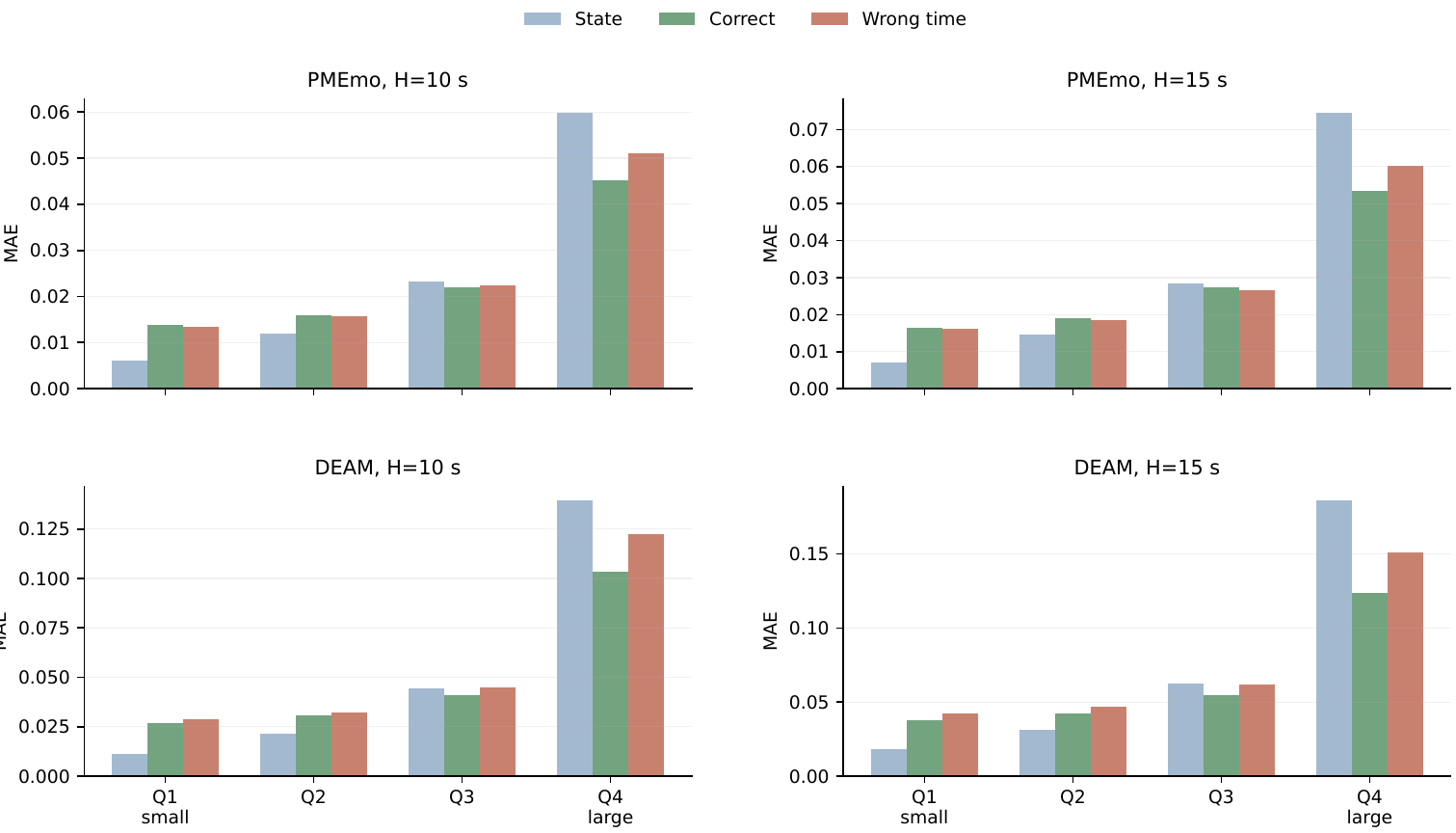}
\caption{MAE stratified by quartiles of absolute true arousal innovation. The state model is difficult to beat for near-zero changes, whereas correctly placed audio provides the largest benefit in Q4. Wrong-time audio retains broad song information but is consistently weaker than the correct future for large changes.}
\label{fig:stratified}
\end{figure*}

CCC measures agreement over the complete innovation distribution, but a practical interpretation also requires knowing where absolute-error reductions occur. Figure~\ref{fig:stratified} divides each dataset/horizon cell into quartiles of absolute ground-truth innovation, with cut points computed within that cell. The state model has the lowest MAE in the smallest-change quartile, where predicting almost no change is an effective strategy. This explains why persistence is a strong local default even when its pooled CCC is low.

The ordering reverses for the largest changes. In every cell, the correctly placed audio representation has lower Q4 MAE than both state and wrong-time audio. For example, DEAM Q4 MAE falls from 0.139 to 0.104 at 10 s and from 0.186 to 0.124 at 15 s; the wrong-time alternatives reach only 0.122 and 0.151. PMEmo shows the same pattern. Thus, scheduled audio earns its aggregate advantage by forecasting substantial departures from inertia, while the near-zero regime remains a boundary where a conservative state prediction is preferable. This analysis was performed on the fixed OOF predictions and does not alter training or selection.

\subsection{Innovation-Label Reliability}
Finally, split-half rater agreement provides scale for the innovation target. Randomly splitting DEAM raters and recomputing two independent group-average arousal trajectories gives CCC 0.472 at 10 s and 0.529 at 15 s over 200 splits (95\% intervals [0.450, 0.493] and [0.504, 0.553]). The complete-representation scores of 0.458/0.564 are in the same range. This comparison is not a human-performance ceiling: each split-half trajectory averages only about half of the raters, whereas the prediction target averages all available raters and is therefore less noisy. A model trained across songs can estimate systematic structure in that all-rater mean and can slightly exceed half-versus-half agreement. This comparison documents disagreement between half panels; it neither bounds attainable performance on the all-rater target nor quantifies the remaining room for improvement.

\section{Discussion}
\label{sec:discussion}

\subsection{Scheduled-Audio Forecast Information}
The 0.389--0.507 CCC lift is the value of the complete scheduled-audio representation over a recent-affect ridge. The future-only analysis shows that much of this information is present in the candidate segment itself: it remains 0.150--0.249 above a nonlinear state model and 0.151--0.261 above past-only audio. This is the primary result.

\subsection{Content Versus Placement}
Song-disjoint downstream fitting keeps held-out songs out of forecasting-head training. Overlap between these songs and the foundation encoders' pretraining corpora has not been excluded. This split also does not remove global musical structure. A wrong-time segment still carries style, production, instrumentation, tempo range, and broad affect. This explains why the complete action lift is larger than the within-song placement margin. The overlapping-shift and acoustic-nearest controls ask a narrower question: whether the correct location remains informative after much of the track-level and local acoustic content is preserved. Their positive margins support temporal specificity, but the residual value of wrong-time audio also shows that the representation uses both track-level content and local placement.

These substitutions are not listener counterfactuals. They test the sensitivity of a fitted forecast while state and song identity are held fixed. A causal music-selection policy would require randomized or logged actions, listener-level outcomes, and assumptions beyond those used here.

\subsection{Relation to Dynamic MER}
Numerical rank claims against ordinary dynamic MER would be misleading. Recognition systems often observe acoustic context at or around the labeled moment, may align labels retrospectively, and may use different song or window splits. Our models expose scheduled audio at a declared origin and score a later innovation. The appropriate comparisons are therefore models trained under the same observation rules. Within that setting, future-only audio remains useful without past descriptors, the complete representation adds further relational structure, and the same-song controls verify temporal specificity. These results provide a reference for the forecasting task, not a replacement for recognition benchmarks.

The same evaluation logic can apply when a future video scene, dialogue response, lesson, or notification is scheduled before its user response is observed: declare the origin, isolate permitted feature support, hold out contextual units, predict change, and test temporal placement with within-context substitutions.

\subsection{Use Case and Limitations}
One immediate use is offline pre-screening rather than automated intervention. Given an observed group-level state and a fixed set of candidate transitions, a media editor could rank or filter segments for human review. The model would estimate expected arousal change; it would not decide which segment causes a desired response. Practical use would additionally require calibrated uncertainty, population matching, listener or group covariates, preference and safety constraints, and prospective validation. Consensus labels cannot represent stable subgroups of listeners who respond in opposite ways.

The study is limited to arousal; the retained analyses do not establish
corresponding results for valence. The models use frozen MuQ representations
and simple regression heads. They do not establish whether a large
fine-tuned sequence model would improve the same forecasting task.

An audit of the earlier explicit-factor experiments found that their
acoustic extraction support could cross the declared observation boundaries.
We therefore exclude those factor predictors and their augmentation,
capacity, backbone, and response-delay results from the evidence reported
in this preprint. They require support-correct feature extraction and
new evaluation before they can support a scientific claim. The retained
predictors use affect state and independently encoded MuQ chunks, without
an explicit-factor correction.

PMEmo and DEAM do not establish cross-cultural, artist-disjoint, or listener-specific transport. The common-song PMEmo analysis removes the immediate 10-s/15-s composition difference, but the horizons still represent different prediction targets and should not be interpreted as a pure response curve. Future work should retain synchronized per-rater trajectories and metadata, group by artist or release, use a denser set of horizons, and evaluate controlled resynthesis that changes one musical factor while preserving others.

Finally, window counts are not independent sample sizes. Song-disjoint fitting and song-cluster resampling address the main dependence, but the reported intervals remain conditional on the fixed representation and development path. They do not replace independent replication.

\subsection{Threats to Validity}
\label{sec:validity}
\textit{Construct validity}: the target is consensus perceived/expressed arousal innovation, not an individual's felt emotion. Split-half agreement indicates label noise, but does not resolve the construct gap.

\textit{Internal validity}: outcome and encoder leakage are addressed through declared support and nesting. Nevertheless, representations can reflect unmeasured production variables, and observational same-song controls cannot establish intervention effects.

\textit{External validity}: results cover PMEmo and an engineering-excluded DEAM evaluation pool at 10 and 15 s. They do not establish performance for other genres, cultures, annotation protocols, or longer policies.

\textit{Statistical conclusion validity}: windows are highly correlated. All splits and bootstrap units are songs, and repeat/fold dispersion is inspected separately. Diagnostic analyses are not relabeled as confirmatory evidence.

\section{Conclusion}
\label{sec:conclusion}
Music may be known before its response is known. This article formalized that distinction as dynamic arousal forecasting with a scheduled future covariate and evaluated it under explicit rules for outcomes, audio support, song splits, and temporal controls. A complete independently encoded MuQ representation improved CCC by 0.389--0.507 over recent affect and remained 0.268--0.382 above a calibrated nonlinear state model, while future-only audio remained well above both that model and a matched past-only trajectory. The first legal 4-s chunk remained informative 5.5--11 s before the target, and correct placement survived strongly overlapping shifts and acoustic-nearest same-song controls. These results show that both scheduled content and its within-song timing contain forecast information.

The contribution is a known-future forecasting framework in which state, future content, and temporal placement are evaluated separately. These results concern consensus arousal under the tested corpora and observation rules, and do not establish causal music-selection effects or generalization to individual listeners.

\section*{Reproducibility and Data Availability}
\label{sec:repro}
All experiments were executed on an Apple M5 Pro system with 24 GB unified memory. Context-isolated MuQ extraction required 5,028 s for 28,495 PMEmo chunks and 2,851 s for 88,409 DEAM chunks; the resulting compressed archives occupy 52 and 161 MB. The frozen MuQ encoder was used only for feature extraction; downstream forecasting fits train PCA and ridge heads. The implementation uses explicit seeds and records source hashes, fold assignments, eligible intervals, encoder revisions, and model settings.

The ancillary materials contain preprocessing and extraction code, forecasting implementations, same-song control construction, grouped-bootstrap utilities, stored aggregate audits, and compressed prediction files. A prediction-recomputation script permits inspection of the principal point estimates and intervals without rerunning the music encoder. The preprint excludes the earlier explicit-factor predictor results after a code audit identified centered and whole-song feature operations that did not satisfy the declared audio-support rule. Legacy factor outputs are not evidence for this preprint's conclusions.

PMEmo and DEAM must be obtained under their original distribution terms~\cite{zhang2018pmemo,aljanaki2017deam}; raw audio and individual participant records are not redistributed. Rebuilding the experiments requires those datasets and the frozen encoder checkpoints. Fixed seeds and stored-prediction agreement support inspection but do not certify full independent reproduction or the absence of all preprocessing errors.

The historical DEAM manifest records the 305 engineering songs and the settings frozen before the complete evaluation. The current implementation's hash differs from the method hash recorded in that manifest. The historical manifest is preserved unchanged, but hash continuity of the current code cannot be established from the supplied files. We therefore treat the date and original code-freeze claim as historical provenance rather than an independently verified guarantee. The evaluation uses repeated song-disjoint outer cross-validation within the 1,494-song pool excluded from engineering development; it is not one permanently isolated outer test fold.

\section*{Acknowledgements}
The authors thank the creators and annotators of PMEmo and DEAM for making their data available. The authors used OpenAI Codex for language editing, document organization, reference and mathematical-consistency review, \LaTeX{} preparation, and software-assisted checks of stored predictions and experimental code. The authors remain responsible for the research, interpretations, and manuscript contents.

\section*{Author contributions}
B.W. designed the study, implemented the method and experiments, analyzed the results, and drafted the manuscript. D.-Q.Z. contributed to the statistical analysis and manuscript review. T.-P.T. supervised the study and reviewed and revised the manuscript. All authors read and approved the final manuscript.

\section*{Funding}
This work received no specific grant from any funding agency in the public, commercial, or not-for-profit sectors.

\section*{Availability of data and materials}
PMEmo and DEAM must be obtained from their original distributors under their respective terms~\cite{zhang2018pmemo,aljanaki2017deam}; source audio and individual participant records are not redistributed. The ancillary materials contain implementation scripts, aggregate audit summaries, and compressed consensus-target prediction files for inspecting the reported music-forecasting results. The reproducibility statement describes their scope and verification limits.

\section*{Declarations}

\subsection*{Competing interests}
The authors declare no competing interests.

\subsection*{Ethics approval and consent to participate}
Not applicable. This secondary analysis collected no new human-subject or animal data; it uses the existing, publicly documented PMEmo and DEAM music-emotion annotation datasets.

\subsection*{Consent for publication}
Not applicable.

\bibliographystyle{unsrtnat}
\bibliography{references}

\end{document}